\documentclass[]{article}

\usepackage[english]{babel}
\usepackage{graphicx}
\usepackage{makecell}
\usepackage[ruled]{algorithm2e}

\usepackage{multirow}

\usepackage{booktabs}
\usepackage{float}

\usepackage{indentfirst}
\title{Sample selection with noise rate estimation in noise learning of medical image analysis}
\author{Maolin Li, Giacomo Tarroni}

\begin{document}

\maketitle

\begin{abstract}
\indent In the field of medical image analysis, deep learning models have demonstrated remarkable success in enhancing diagnostic accuracy and efficiency. However, the reliability of these models is heavily dependent on the quality of training data, and the existence of label noise —errors in dataset annotations— of medical image data presents a significant challenge. 
\\
\indent This paper introduces a new sample selection method that enhances the performance of neural networks when trained on noisy datasets. Our approach features estimating the noise rate of a dataset by analyzing the distribution of loss values using Linear Regression. Samples are then ranked according to their loss values, and potentially noisy samples are excluded from the dataset. Additionally, we employ sparse regularization to further enhance the noise robustness of our model.
\\
\indent Our proposed method is evaluated on five benchmark datasets and a real-life noisy medical image dataset. Notably, two of these datasets contain 3D medical images. The results of our experiments show that our method outperforms existing noise-robust learning methods, particularly in scenarios with high noise retes.
\end{abstract}

%1 Introduction
\section{Introduction}
\indent Deep learning has been widely used in medical image analysis tasks and achieved remarkable success. The application of various neural network architectures has enabled more precise identification of pathologies from all kinds of medical imaging modalities, such as X-ray, CT scans, MRI, ultrasound, and so on. These advancements have greatly accelerated and refined the decision-making process in clinical environments. 
\\
\indent Despite the current success, challenges still exist with label noise emerging as a notable issue. ‘Label noise’ refers to the incorrect labels caused by various reasons, especially by the mistake of annotators. In deep learning-based image analysis tasks, the availability of accurately labeled images is imperative for the neural network to learn effectively and enhance its performance. Noisy labels hurt this process because neural network models might overfit to label noise. 
\\
\indent Label noise is a severe problem in the domain of medical image analysis, mainly for two reasons. The first reason is the difficulty and time consumption to label medical images. Analysing medical images requires expertise of medical imaging knowledge and experience, which is expensive to acquire. Furthermore, the privacy constraints of patient data made it difficult to collect large amounts of medical image datasets. The second reason is the inherent diagnostic limitations of medical images. Some types of medical images naturally can not provide an absolutely precise representation of the actual physiological conditions within the human body. Consequently, even if the labeller could do a perfect job, the presence of label noise in medical image labels remains unavoidable. Additionally, severe inter-observer variability among experts \cite{1bridge2016intraobserver}\cite{2karimi2019deep} further compounds this issue, which means that top experts could have different interpretations of the same image and the potential for different labels.
\\
\indent Many studies have proved that noisy labels have negative impact on the performance of neural network models \cite{3krause2016unreasonable}\cite{4arpit2017closer}. Various noise-robust learning methods have been proposed to address the issue of label noise. However, most of the existing noise-robust learning techniques still have not been applied to medical images \cite{5karimi2020deep}. In this project, we aim to explore the efficacy of noise-robust learning approaches in medical images and propose our original method.
Our key contributions are summarized as follows:
\\
\indent i) We have introduced a three-phase learning scheme to filter clean data from a noisy medical image dataset. We have also incorporated GCE loss function and sparse regularization to further enhance its robustness.
\\
\indent ii) We have proposed a noise rate estimation module based on loss value distribution and proved its efficacy. Unlike previous works, our proposed noise rate estimation module is based on linear regression. 
\\
\indent iii) We have implemented and compared our original method with various existing noise-robust learning methods in medical images. Experimental results prove that our proposed method achieves superior performance in different kinds of medical images.
\\
\indent Specifically, to the best of our knowledge, our study stands as the first to validate the effectiveness of noise-robust deep learning algorithms on 3D medical images like CT and MRI.

%2 Related work
\section{Related work}

% general noise-robust learning methods
\subsection{Noise-robust learning for natural images}

\indent Some early studies modified the architecture of neural networks to make them work better on noisy datasets, mostly by adding noise-adaption layers. For example, Sukhbaatar et al. \cite{sukhbaatar2014noiseadapationlayer} added an extra noise layer, which adapts the network outputs to match the distribution of label noise. 

\indent Some other studies implemented noise-robust loss function or regularization methods specifically designed towards noisy datasets. Wang et al. \cite{sce} noticed that cross entropy loss easily overfits the false labels but underfits hard correct labels. Inspired by symmetricity of Kullback-Leibler divergence, they combined CE loss with Reverse Cross Entropy and proposed a more noise-robust loss function called Symmetric Cross Entropy. MixUp \cite{mixup} is a commonly used data augmentation method that prevents overfitting. New data is synthesized simply by the linear interpolation of two training examples randomly chosen from the training dataset.

\indent Noisy labels lead to inaccurate loss values, so correcting the labels and loss values is another idea to achieve noise robustness. Some studies tried to build a transition matrix to correct the loss values. Patrini et al. \cite{patrini2017} proposed two procedures for estimating the transition matrix and correcting the loss, including ‘forward correction’ and ‘backward correction’. These two procedures are converse with each other but will improve the robustness to noise labels. Some other studies ignored the possibly noisy labels and transpose the noise learning problem to a semi-supervised learning problem. For example, DivideMix \cite{dividemix} uses two-component and one-dimensional Gaussian mixture models to transform noisy data into labeled (clean) and unlabeled (noisy) sets. Then, it applies a semi-supervised technique MixMatch \cite{mixmatch}.

\indent Another straightforward approach against label noise is finding the incorrectly labelled samples and removing them from the dataset. Decoupling \cite{6malach2017decoupling}, an early strategy in this domain, involves maintaining two networks simultaneously. For every mini-batch, only the samples that receive different predictions from the two networks are used to update the neural network parameters. This strategy is often referred to as ‘update by disagreement’. MentorNet \cite{7jiang2018mentornet} uses a mentor network and a student network. The mentor network identifies small-loss samples and guides the training of student network by only feeding small-loss samples which are likely to be correctly labelled. Co-teaching \cite{8han2018co} and Co-teaching+ \cite{9yu2019does} both maintain two networks. In Co-teaching, The two networks  select samples with minimal losses and then feed them to its peer network. Co-teaching+ further integrated the ‘update by disagreement’ strategy from DeCoupling into Coteaching, thereby combining the strengths of both approaches. Some other studies only maintain one network, but implement multiple training rounds or phases to select clean data. In the method proposed by Shen et al. \cite{11shen2019learning}, during each training round, the model removes high-loss samples and the remaining small-loss samples are left for training in the next round. O2U-Net \cite{o2u} incorporates three training stages: initial pre-training, followed by a second stage where this model identifies and removes high-loss samples to refine the dataset, and a final stage where training restarts with the cleaned dataset. MORPH \cite{13song2021robust} shares a similar idea with O2U, but is able to switch its learning phase at the transition point automatically. It also introduces the concept of memorized examples. Obviously, A direct method to mitigate label noise involves identifying and removing inaccurately labeled samples from the dataset. Decoupling, an early strategy in this domain, employs dual neural networks to filter potentially clean data. This approach involves maintaining two networks that operate simultaneously. For each mini-batch, only the samples receiving divergent predictions from the two networks are used to update the neural network parameters, a strategy often termed 'update by disagreement.' 

MentorNet \cite{7jiang2018mentornet} introduced a mentor-student network arrangement, where the mentor network identifies small-loss samples and guides the training of the student network by selectively providing these likely correctly labeled samples. Similarly, Co-teaching \cite{8han2018co} and Co-teaching+ \cite{9yu2019does} utilize two networks. In Co-teaching, both networks select samples with minimal losses and exchange these for mutual training. Co-teaching+ integrates the 'update by disagreement' approach from Decoupling into Co-teaching, amalgamating the benefits of both methodologies.

Other studies employ a single network but implement multiple training rounds or phases to filter clean data. In the approach by Shen et al. \cite{11shen2019learning}, high-loss samples are removed in each training cycle, while only the remaining small-loss samples are utilized for subsequent training rounds. O2U-Net \cite{o2u} incorporates three training stages: an initial pre-training, followed by a second stage where this model identifies and eliminates high-loss samples to refine the dataset, and a final stage where training recommences with the cleaned dataset. MORPH \cite{13song2021robust}, sharing similarities with O2U-Net, can autonomously switch its learning phase at a transition point and introduces the concept of memorized examples.

%noise robust learning in medical image analysis
\subsection{Noise robust learning for medical images}
\indent The study of Dgani et al. \cite{16dgani2018training} was one of the earliest to apply noise robust deep learning techniques to medical images. They added a noise-robust layer to a neural network in a mammography classification task and slightly improved the accuracy score. Pham et al. \cite{17pham2021interpreting} implemented label smoothing techniques in the classification of chest X-ray images. By comparing their method with other noise-robust methods, such as ignoring noisy samples, they proved that label smoothing could enhance the AUC (area under ROC curve curve) by up to 0.08. Xue et al. \cite{18xue2019robust} presented a two-stage strategy for learning from corrupted skin lesion datasets. The first stage involved uncertainty sample mining to eliminate the noisy-labelled data, and the second stage employed a data re-weighting method. This approach improved the classification accuracy by 2\%-10\%, depending on the noise level.
\\
\indent Co-correct \cite{19liu2021co} implemented a dual-network model to filter potentially noisy data and correct them. Unlike some other dual-network strategies such as co-teaching, Co-correct calculates loss values for potentially noisy samples but sets these values to zero. This model was tested on two histopathology datasets: ISIC-Archive (skin melanoma histopathology) and PatchCamelyon (lymph node histopathology). The findings proved Co-correcting outperformed comparative models in accuracy. Xue et al. \cite{20xue2022robust} employed a two-step approach. The first step involved the selection of clean samples, followed by collaborative training in the second step. This strategy was also found to be effective in the classification of pathological slides. 
\\
\indent Hu et al. \cite{21hu2023fundus} developed a mixed noise-robust method for classifying fundus images. They first applied data cleansing strategy to filter out noisy data based on confidence of prediction. Then, an adaptive negative learning model was employed to modify the loss function, complemented by sharpness-aware minimization to fine-tune both loss and sharpness. Zhu et al. \cite{22zhu2021hard} introduced a hard-sample-aware approach designed for learning from noisily labelled histopathology images. They used a detection model to identify easy, hard and possibly noisy samples, thereby creating a cleaner dataset. By using a noise suppression and hard-enhancing method, training on the refined dataset obtained better results when tested on DigestPath2019, Cemelyon16 and Chaoyang dataset. Khanal et al. \cite{23khanal2023improving} recognized the efficacy of self-supervised pre-trained weights on noisy natural image datasets and extended this approach to medical datasets (NCT-CRC-HE-100K histological image dataset and COVID-QU-Ex X-ray dataset) and demonstrated its effectiveness. Jiang et al. \cite{24jiang2023label} integrated contrastive learning and intra-group attention mixup strategies in their methodology. This method was tested on three medical image datasets: Retina OCT, Blood Cell and Colon Pathology images and exhibited relatively strong performance. Zhu et al. \cite{25zhu2023robust} combined two modules: a noise rate estimation module and a noisy label correction module. Evaluation on ISIC-2016 skin pathology dataset and an original ultrasound image dataset indicated superior performance relative to other noisy learning methods. Chakravarthi et al. \cite{26chakravarthi2023sparsely} proposed a sparsely supervised learning strategy based on transfer learning and applied it to the classification of skin cancer images from ISIC dataset, showcasing its applicability and effectiveness in this domain.

%3 Methods
\section{Methods}

\subsection{Problem setting}
\indent Consider a medical image dataset, denoted as $D$, has $n$ images and corresponding labels, i.e.
$$
D\ =\ {(x_i,\ \hat{y}_i)\ |\ 1\le\ i\le\ n}
$$
\indent For a sample $x_i$ , if its annotated label $\hat{y}_i$ matches the true indication of the medical image (the correct label $y_i$), we call it a clean sample. Otherwise, it is a noisy sample. Here we let $\eta$ represent the noise rate, a parameter which should be unknown to the neural network model. If this dataset has $k$ classes, the noise rate $\eta$ should be smaller than $\frac{k-1}{k}$.
\\
\indent Our primary objective is to find a mapping function $f: x \rightarrow y$, where $x$ is the above mentioned medical image, and $y$ is the true label. The function $f$ describes the complex relationship between the image and its corresponding label. Specifically, in this project, $f$ is modeled by a deep neural network ending with a SoftMax layer.
\\
\indent In our problem setting, the ground truth label $y$ for a given sample is unknown due to various reasons, such as misdiagnosis or disagreements between labellers. This means that during the neural network training, we only have $\hat{y}$ assigned for each sample, which can possibly be incorrect with rate $\eta$. We have to train this network with the image $x$ and its annotated label $\hat{y}$ given the unavailability of true label ${y}$. It has been reported that none-noise-robust training strategy with noisy labels can lead to degradation of accuracy on test set. Our aim is to find a solution to optimize the neural network classifier $f$ with ${x,\ \hat{y}}$ to achieve comparable results with a model trained on clean data ${x,\ y}$.
\\
\indent In the following sections, we will present our original training strategy to address the challenges posed by label noise. Compared with a none-robust baseline approach, our method is composed mainly of three modules: A noise-rate estimation module based on the distribution of loss values, a three-stage training scheme to select clean data, and sparse regularization with output permutation to further enhance noise robustness.

%3.1 Noise rate estimation with linear regression
\subsection{Noise rate estimation with linear regression}
\indent A ideal noise-robust learning methods that employ sample selection strategy need to know how much data it should forget or remember. This revolves around a parameter commonly referred to as "forget rate" or "remember rate", which theoretically should be close to the actual noise rate, $\eta$, of the dataset. This principle is intuitive: if the noise rate is significantly smaller than forget rate, some clean samples will be erroneously deleted, resulting in a waste of data. Conversely, if the noise rate is significantly larger than the forget rate, some noisy samples may remain unfiltered and thus still be left in the dataset.
\\
\indent However, for a real-life medical image dataset, obtaining its precise noise rate is often impossible. In this section, we introduce a noise rate estimation module based on the distribution of Cross-Entropy loss value across all training samples. Auxiliary datasets are incorporated to better explore the distribution pattern of loss values under different noise rates. Specifically, we utilize five medical image datasets as auxiliary datasets in this study, enabling our method to work independently of any prior knowledge about the primary dataset under investigation.
\\
\indent Our implementation of the noise rate estimation module begins with randomly corrupt the labels of some samples in the auxiliary datasets according to predefined noise ratios. Then, we perform none-robust baseline deep learning training on these corrupted datasets, recording the distribution of Cross-Entropy loss values. Next, a Linear Regression model is employed to learn from these distribution patterns. These training steps are conducted with only auxiliary datasets before we explore our target dataset. The following algorithms details this process.

%Algorithm table of linear regression
\begin{algorithm}
	\KwData{Linear regression model $L$, Auxiliary datasets $D_1, D_2, D_3, ... D_i$}
	
	\ForEach{\rm{Auxiliary dataset $D_i$}}
	{Add label noise with noise rate $\eta$ to $D_i$ and get noisy dataset $D{'}_i$ \;
		Do normal neural network training on $D{'}_i$\; 
		Obtain loss value distribution $R$, number of classes $c$, number of samples $N$}
	
	Train $L$\ with loss distribution $R$, number of classes $c$, number of samples $N$, noise rate $\eta$ \;
	\KwResult{A linear regression model $L$ which can predict noise rate}
	\caption{Train a noise-rate estimator with linear regression}
\end{algorithm}

\indent Training this Linear Regression model involves three inputs: the distribution of Cross-Entropy loss values, the total number of samples, and number of classes in the auxiliary dataset. In detail, after recording all the loss values, they are organized in descending order and divided into $j$ intervals. We will count the number of samples within each interval and calculate their respective ratios. These ratios, combined with the total number of samples $N$ and number of classes $c$ of the dataset, will be used as inputs of Linear Regression model $L$ for fitting the noise rate $\eta$, which can be denoted by: 

$$
\eta=\sum_{i=1}^{j}{\left(k_i\frac{n_i}{N}\right)\ +\ k_{j+1}N\ +\ k_{j+2}c + b}
$$

\indent Where $\eta$ represents the noise rate of the dataset that we want to predict, while $k$ and $b$ are parameters that define the linear regression model. Additionally, $n_i$ denotes the number of samples whose loss values fall with in the $i-th$ interval. In this study, we set the value of j as 1000, which allows for a more precise estimation of the noise rate of a dataset based on the loss value distribution of samples.

%3.2 Data selection with three-phase training scheme
\subsection{Data selection with three-phase training scheme}

\indent In this section, we employ a three-phase training scheme to filter out noisy data.
\\
\indent Pre-training: In this phase, we pre-train the network directly on the original dataset, inclusive of potentially noisy labels. The aim of this phase is just to lay the foundation for the second phase.
\\
\indent Data filtering: In this phase, we calculate the cross-entropy loss values of all samples and rank them in descending order. Next, leveraging the noise-rate estimator that we have implemented, we estimate the noise rate of the dataset based on the loss value distribution. It is important to note that our forget rate does not precisely equal the predicted noise rate. Ideally, forget rate should equal noise rate, ensuring the deletion of noisy data is thorough and only clean data is left. However, given the uncertainty in data filtering accuracy, our implementation sets the forget rate slightly smaller than the predicted noise rate. This allows for the preservation of more data in the cleansed dataset, with additional noise-robust methods available to further enhance robustness.
\\
\indent Training on Clean Data: In the last phase, we re-initialize the parameters of the network, and conduct final training by combining another noise-robust regularization strategy on the cleansed dataset.
The following algorithm presents the whole training process of our proposed method.

%Algorithm table of whole process
\begin{algorithm}
	\KwData{Dataset $D$ including a fraction of noisy labels.}
	
	\textbf{Stage1: Pre-training} \\
	\textbf{Initialization:} Parameters $W$ of deep neural network classifier $f$
	\Repeat{\rm{Accuracy and loss stable or reach maximum epoch number}}
	{Do ordinary image classifier training on $D$}
	Save $W$, obtain loss value of every sample $l_n$
	
	\textbf{Stage 2: Data selection}
	
	Estimate noise rate $\eta$, forget rate $k$ with linear regression model $L$ \\
	Obtain $R$ by ranking all the samples in descending order according to $l_n$ \\
	Remove top $k\%$ samples from $D$ to obtain a cleaned dataset $D’$
	
	\textbf{Stage 3: Training on cleaned data}
	Load pre-trained weights $W$ \\
	\Repeat{\rm{Accuracy and loss stable or reach maximum epoch number}}
	{Do ordinary image classifier training on $D'$}
	
	\KwResult{Obtain the image classifier $f$} 
	\caption{Train a noise-rate estimator with linear regression}
\end{algorithm}

%3.3 Noise-robust sparse regularization with output sharpening
\subsection{Noise-robust sparse regularization}

\indent In this section we introduce the sparse regularization strategy employed in the third phase of our three-stage training scheme to train the cleansed dataset. Despite the removal of some noisy data in the second phase, the cleanliness of remaining samples remains uncertain. Therefore, we use this regularization strategy to further improve the robustness in the last training stage. Notably, this strategy also has potential benefits if implemented in the first pre-training phase. Ideas in this section are mostly learned from \cite{lnlsr}.
\\
\indent It has been proved that restricting the output of neural networks to a one-hot form will grant robustness to any loss function, and that when combined with $l_p$ norm regularization, this method will improve the performance under noisy datasets to a higher level \cite{lnlsr}. Our sparse regularization strategy composes mainly of three parts: noise robust GCE loss function, network output sharpening, and $l_p$ norm regularization. 
\\
%GCE Loss function
\indent \textbf{Generalized cross entropy (GCE) loss function:} This is a synthesized approach, working as a middle ground between Cross Entropy (CE) loss and Mean Absolute Error (MAE) loss. It combines the noise robustness from MAE loss and the convergence from CE loss. Mathematically denoted as:
$$
L_q(f(x),\ y_j)=\frac{1-f_j{(x)}^q}{q}
$$
\indent Where $f(x)$ presents the output of the neural network and $y_j$ denotes the label. Moreover, $q$ is a parameter between 0 and 1, determining the compromise between the two loss components. When $q$=1, this will be MAE loss and when $q$ approaches 0, it will become Cross Entropy loss. 
\\
%output permutation
\indent \textbf{Output Permutation:} The purpose of the output permutation module is to transform the network output to resemble a one-hot vector. One popular strategy to approximate a one-hot vector by the continuous mapping is to use a temperature-dependent SoftMax function, expressed as:
$$
\sigma_\tau{(z)}_i=\frac{\exp({z}_i/\tau)}{\sum_{j=1}^{k}\exp({z}_j/\tau)}
$$
\indent Here $z$ represents the output of the neural network, and $\tau$ is a parameter referred to as temperature. When $\tau$ is smaller (or we say when the ‘temperature’ is lower), the output will converge to more like a one-hot vector.
\\
%Sparse regularization
\indent \textbf{$L_p$ Norm Regularization:} We further employ $l_p$ norm regularization to promote the sparsity of network output. The $l_p$ norm regularization can represented as:
$$
\lambda{||f(x_i)||}_p^p
$$
\indent Here $f(x_i)$ denotes the output of the neural network after SoftMax layer. The parameter $p$ is a parameter between 0 and 1 that controls the strength of regularization. After adding the regularized value, the final loss value which the neural network aims to minimize will be:
$$
\sum_{i=1}^{n}{L_q\left(f\left(x_i\right),\hat{y}\right)+\ \lambda||f(x_i{)||}_p^p}
$$
\indent Where $n$ denotes the size of the dataset, $L_q$ refers to the GCE loss function, $f$ is denotes the neural network image classifier and $\hat{y}$ is an annotated label.

%Experiments
\section{Experiments}
\indent This section assesses our proposed method on six medical image classification datasets. An additional ablation study is conducted to validate the efficacy of our original noise-rate estimation module. Comparisons with multiple state-of-the-art noisy learning methods are elaborated in this section.
\\
\indent

%4.1 Datasets
\subsection{Datasets}

\indent \textbf{Medmnist} \cite{medmnist}: MedMNIST is a collection of MNIST-like standardized biomedical images. This collection includes various types of medical images, all pre-processed into 28x28 (2D) or 28x28x28 (3D) with the corresponding classification labels. Three 2D (PathMNIST, OCTMNIST, PneumoniaMNIST) and two 3D (OrganMNIST3D, VesselMNIST3D) datasets from MedMNIST are employed in this paper.

%Path
\indent PathMNIST \cite{28kather2019predicting}: This dataset comprises pathological sections routinely collected from colorectal cancer operations, undergoing haematoxylin and eosin staining. This task involves classifying the nine pathological subtypes of colorectal cancer histology. This dataset contains 107,180 samples in total (89,996 for training, 10,004 for validation, 7,180 for testing). The test dataset is provided by a distinct medical centre, ensuring diversity from the training and validation sets.

\begin{figure}[H]
	\centering
	\includegraphics[scale=1.0]{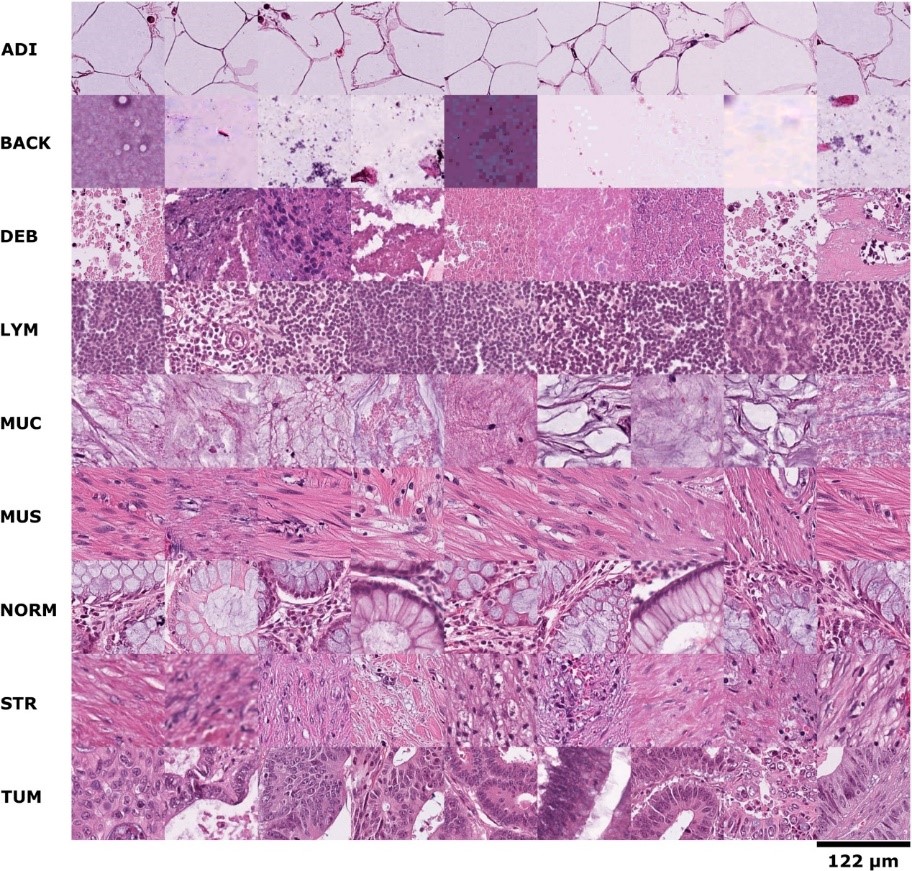}
	\caption{Sample images in PathMNIST}
	\label{path}
\end{figure}

%OCT
\indent OCTMNIST \cite{29kermany2018identifying}: This dataset composes of eye coherence tomography (OCT) images for diagnosing retinal diseases. It contains 109,309 samples (97477 for training, 10832 for validation, 1000 for testing) with four types of OCT images: Normal, CNV (Choroidal Neovascularization), DME (Diabetic Macular Edema) and Drusen (some kinds of desposits beneath retina). All the images from this dataset are naturally in grey scale.

\begin{figure}[H]
	\centering
	\includegraphics[scale=0.8]{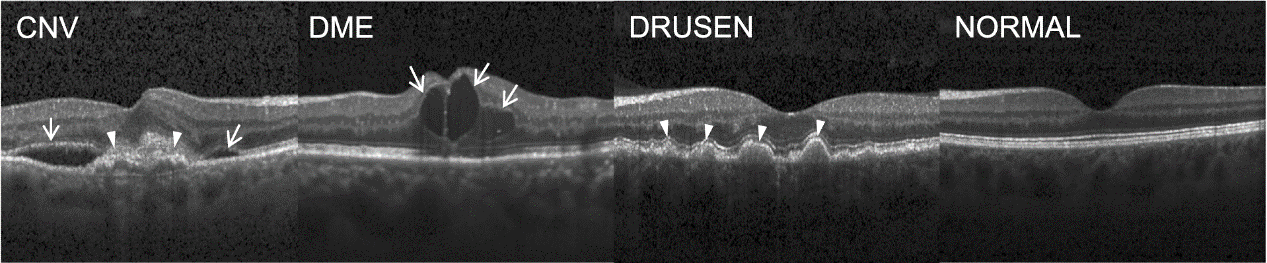}
	\caption{Sample images in OCTMNIST}
	\label{oct}
\end{figure}

%Pneumonia
\indent PneumoniaMNIST \cite{29kermany2018identifying}: A binary-classification dataset for diagnosing pneumonia through chest X-ray films. The original dataset includes 5856 cases (5232 for training and 624 for testing) and the training set was further split into training and validation with a 9:1 ratio. 

%Organ
\indent OrganMNIST3D \cite{30bilic2023liver}\cite{31xu2019efficient}: Comprising abdominal CT images, this dataset targets the classification of abdominal CT organs. This dataset contains 1,743 CT images (972 for training, 161 for validation, 610 for testing). 3D bounding boxes that contains the target organs are extracted from the raw CT images and resized to the same size for multi-class classification of 11 abdominal organs.

\begin{figure}[H]
	\centering
	\includegraphics[scale=2.4]{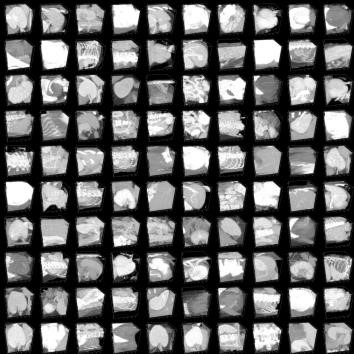}
	\caption{Sample images in OrganMNIST}
	\label{organ}
\end{figure}

%Vessel
\indent VesselMNIST3D \cite{32yang2020intra}: This dataset is based on an open-access intracranial aneurysm dataset. This is a binary-class classification task for diagnosing the presence of aneurysm from cranial Magnetic Resonance Angiography. This dataset contains 1,694 normal artery segments and 215 aneurysm segments. 

\begin{figure}[H]
	\centering
	\includegraphics[scale=0.7]{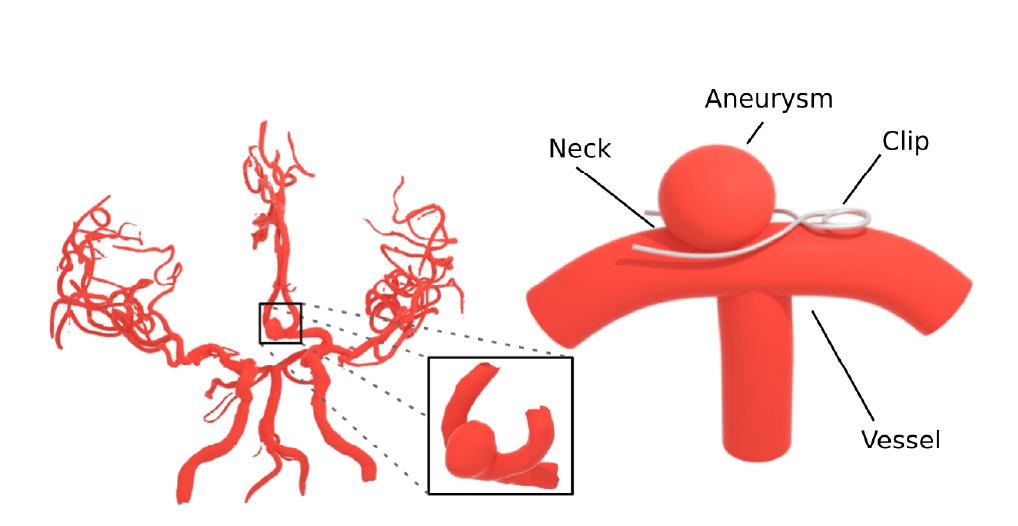}
	\caption{Sample images in VesselMNIST}
	\label{vessel}
\end{figure}

%Noise implementation
\indent For above five MedMNIST datasets, original labels are treated as ground truth, assuming that they are all correctly labelled with no noise. To prove the robustness in learning with noisy labels, we introduce artificial noise to these data by corrupting some of the labels. For all of these five datasets, we implemented symmetric noise by randomly flipping the labels to any other class under noise rates. Additionally, for PathMNIST and OrganMNIST which are both multiple classification tasks, we further implemented asymmetric noise by flipping the labels only to its next class under the noise rate, as described in \cite{odnlasymmetricmethod} and \cite{24jiang2023label}. We evaluate the performance of different methods across five fixed noise rates: 0, 0.1, 0.2, 0.3, and 0.4. Artificial label noise is added only to the training set, so the validation and test set are still regarded as clean.

%NoisyCXR
\indent \textbf{NoisyCXR}: NoisyCXR is a binary classification task aimed at identifying the presence of pneumonia in chest X-ray images. This dataset was developed to emulate the real-world label noise in medical images, instead of adding artificial noise to clean datasets. Specifically, Our implementation details followed the methodologies described in \cite{noisycxrmicrosoft}. The X ray images in NoisyCXR are all provided by National Institutes of Health (NIH), with original labels (termed ‘NIH labels’) that are generated using a language-processing approach based on radiological reports. These labels encompass 14 classes representing various abnormalities, with one single X-ray potentially tagged with multiple NIH labels. However, this automated labeling process was proved to introduce certain degree of label noise. The labels were then manually validated and reissued by Radiological Society of North America (termed RSNA labels), which are regarded as clean labels in our study. The RNSA labels include three groups: normal, pneumonia-like opacity, and other abnormalities. For the construction of the NoisyCXR dataset's clean labels, cases annotated with 'Pneumonia-like opacity' are classified as 'pneumonia', and all others are designated as 'no pneumonia'. 
\\
\indent We utilized the NIH labels to generate the noisy labels of NoisyCXR. In the case of images labeled as 'Pneumonia' in the NIH labels, they retain their classification. However, there are many X-ray images with ambiguous terms such as 'consolidation' or 'infiltration', which may suggest potential pneumonia but lack definitive confirmation. In order to simulate how label noise really happens in a clinical situation, we attributed all cases with label 'consolidation' or 'infiltration' to class 'pneumonia'. For all other labels, encompassing both 'no abnormality' and any other types of abnormalities, they are categorized under the 'no pneumonia' class. Through this rigorous process, we have crafted the NoisyCXR dataset, with a noise rate of 40\%, thereby providing a robust framework for evaluating the performance of binary classification models in the presence of label noise.
\\
\indent

\begin{figure}[H]
	\centering
	\includegraphics[scale=0.83]{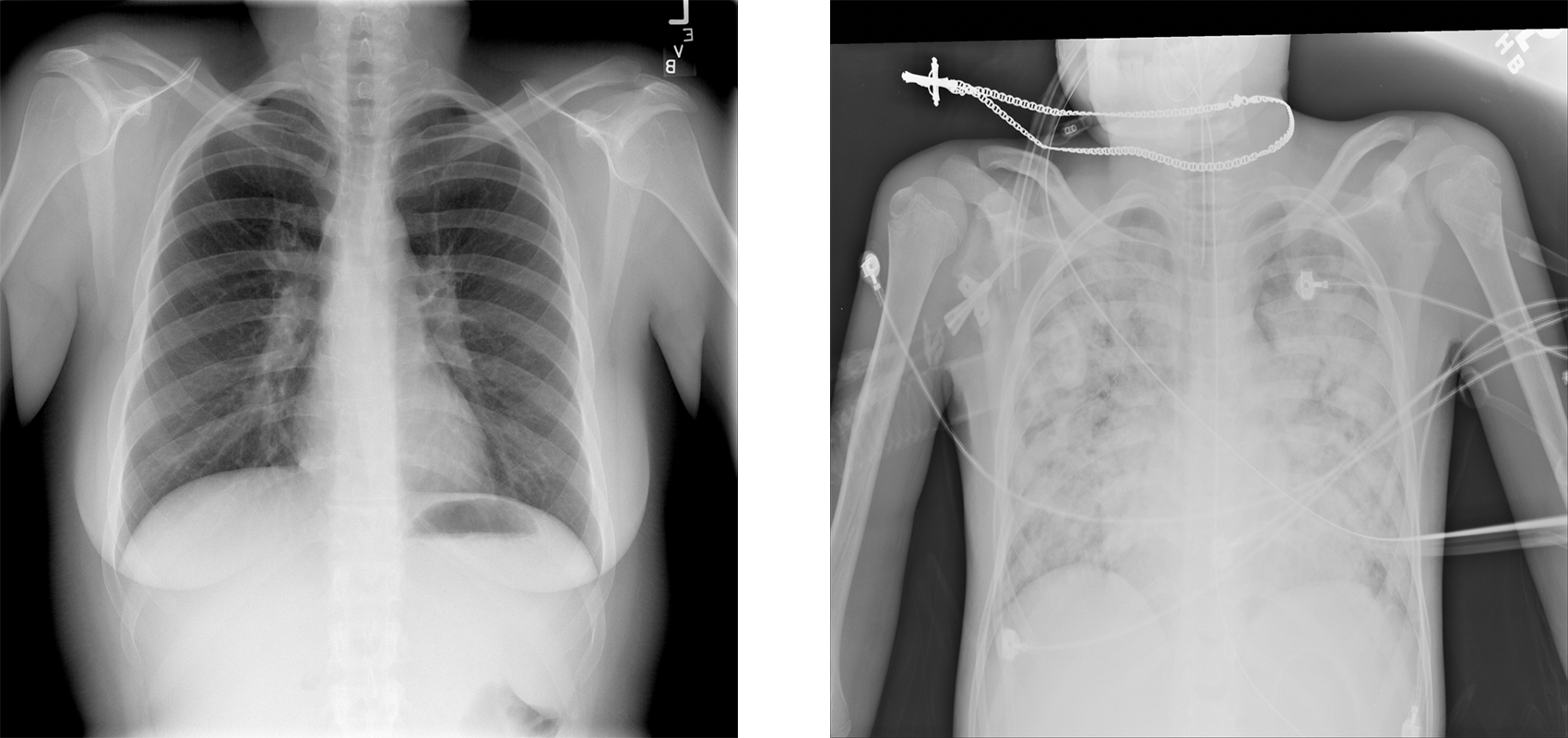}
	\caption{Sample images in NoisyCXR dataset: no pneumonia / pneumonia}
	\label{noisycxrsampleimage}
\end{figure}

%4.2 Benchmark comparison methods
\subsection{Benchmark comparison methods}

\indent To evaluate the effectiveness of our proposed method, we implemented a none-robust baseline model and several state-of-the-art noise robust training methods for comparison. Two of them (O2U-Net and Coteaching\_plus) employed sample selection methods to filter data, and the sparse regularization method that we used was proposed by LNL\_SR.
\\
\indent Baseline: The baseline method in our experiment is a common training procedure without using any noise-robust strategies. It serves as a simple benchmark only for comparative purposes.
\\
\indent O2U-Net \cite{o2u}: This method uses three-phase training to remove the possibly noisy data by ranking the loss values of all samples. More importantly, noticing the difficulty to judge whether the network is being overfitting or underfitting, the authors employed Cyclical Training to calculate the mean loss value through multiple epochs before filtering high-loss samples. 
\\
\indent MixUp \cite{mixup}: Mixup is a well-known data augmentation method, featuring simple but effective linear interpolation. This method is intended for avoiding overfitting in clean datasets, but has been proved useful on some noisy datasets as well.  
\\
\indent Coteaching plus \cite{9yu2019does}: Coteaching plus maintains two networks: each network selects small-loss samples from its training data and feeds them to the other network to learn from. In this way, high-loss samples in each mini-batch are treated as noisy and ignored. Network parameters are updated based on samples where the networks disagree with each other.
\\
\indent CDR \cite{cdr}: This is an early-stopping method designed to prevent the network from overfitting to noisy data. The authors categorize neural network parameters as important and none-important, employing early-stopping on the parameters that are more likely to cause overfitting.
\\
\indent Self-adaptive Training \cite{selfadaptive}: This is a label refurbishment method that corrects the labels by combining original labels with predictions and applying an exponential moving average strategy to stabilize the occasionally unreliable outputs of neural networks.
\\
\indent Multiclass \cite{multiclass}: This method employs a two-stage loss reweighting strategy to minimize the impact of incorrectly labeled cases. In the first stage, the model is pre-trained to calculate a weight transfusion matrix, which is then used in the second stage to estimate the true loss value of each sample.
\\
\indent LNL\_SR \cite{lnlsr}: LNL\_SR proves that any loss function can be made robust to noisy labels by restricting the network output to the set of permutations over a fixed vector. This method applies network output sharpening with a tempered SoftMax function, and $L_p$-norm regularization to promote network outputs to be sparse. The sparse regularization approach is combined with GCE loss function to further enhance its noise robustness.

%4.3 Implementation Details
\subsection{Implementation Details}

\indent We selected ResNet-18 as the backbone network for all experiments. The training batch size was set as 128 and training epochs were fixed as 200, aligning with the default settings in most comparison methods. The remaining hyperparameters of existing comparison methods are all retained in accordance with the original code. Two of the seven implemented comparative noise-robust methods (O2U-Net and Coteaching\_Plus) require a forget rate to know how much data to filter, and in this experiment we set the forget rate to at constantly 0.2. These choices ensured the most fairness between different methods.

\indent For our proposed method, the first and third phases were allocated 90 epochs each, with the remaining 20 epochs left for the second phase. Following LNL\_SR, SGD optimizer with learning rate of 0.01 was applied in the first and third phases. While in the second phase, following the implementation details in O2U-Net, a smaller batch size of 16 was implemented and trained by a vanilla ResNet-18 and cross-entropy loss, with learning rate 0.01.

%4.4 Evaluation metrics
\subsection{Evaluation metrics}

\indent In the context, we employ classification accuracy as the main evaluation metric to different methods under various noise rates and noise types. Additionally, in the appendix, we provide other evaluation metrics including F1 scores, ROC curves and AUC score, as well as training curves to further validate our model.

%4.5 Experiments on 2D benchmard datasets
\subsection{Experiments on 2D benchmark datasets}

\indent This section introduces the overall comparisons on three 2D medical image datasets: PathMNIST, OCTMNIST (multi-classification) and PenumoniaMNIST (binary classification). For all of these three datasets, we introduced symmetric noise, and specifically for PathMNIST, we also asymmetric noise. We will report the average accuracy score of the last epoch on the test set through three runs with standard deviation . 

\indent Firstly, from table \ref{accpaths} to table \ref{accpneumonia}, it is evident that both symmetric and asymmetric label noise leads to poor generalization on the test dataset. Additionally, the rate of degradation in accuracy varies greatly according to the specific dataset. Although a few studies \cite{vanillanetworkrobust} have suggested that vanilla neural networks may exhibit a degree of noise robustness even without the application of any noise-robust strategies, our experimental results, aligning with a broader spectrum of research findings \cite{rog}\cite{cdr}\cite{lnlsr}, demonstrate a sharp decrease in testing accuracy with increasing noise rates for both symmetric and asymmetric noise. Evidently, it is important to use noise robust methods on these dataset to improve the classification performance of deep neural networks on such noisy medical image data.

%acc table PathMNIST
\begin{table}[H]
	\centering
	\caption{Average accuracy (\%, 3 runs, with standard deviation) of different methods on PathMNIST dataset with symmetric noise}
	\begin{tabular}{|c|c|c|c|c|c|}
		\hline
		Noise rate 	& 0 & 0.1 	& 0.2 & 0.3 & 0.4 	\\ \hline
		Baseline & 88.96±2.06 & 86.01±0.58 & 78.85±1.22 	& 70.08±0.25 			& 63.60±1.87 			\\ \hline
		O2U   	& \textbf{91.21±0.46} 	& 90.75±0.74 & \textbf{89.38±0.94} 	& 87.31±0.27 & 80.88±0.84 \\ \hline
		MixUp 	& 87.61±0.83 & 85.95±1.05 & 83.60±1.67 			& 77.68±1.91 			& 68.59±1.14 			\\ \hline
		Coteaching+ 	& 89.35±1.48 	& \textbf{90.49±0.32} 	& 89.00±0.34 	& 84.70±1.02 	& 84.42±0.69 	\\ \hline
		CDR   		& 90.79±0.32 & 87.64±1.03 	& 84.02±0.30 			& 77.10±2.70 			& 67.59±1.60 			\\ \hline
		Self-adaptive  	& 90.06±1.79 & 87.89±0.77 	& 84.05±0.57 	& 78.71±0.76 			& 68.93±1.25 			\\ \hline
		Multiclass 	& 89.47±1.30 	& 85.39±1.02 & 80.86±0.56 	& 71.62±0.98 			& 61.61±0.49 			\\ \hline
		Lnl\_sr 		& 88.15±1.47 		& 87.87±1.51 	& 86.03±1.06 	& 87.07±0.71 	& 86.26±1.11 			\\ \hline
		Ours  	& 89.98±1.45 		& 87.5±0.98 	& 88.20±1.86 	& \textbf{88.73±0.98} 	& \textbf{87.76±0.47} 	\\ \hline
	\end{tabular}%
	\label{accpaths}%
\end{table}%

 % acc table PathMNIST assymtric
 \begin{table}[H]
 	\centering
 	\caption{Average accuracy (\%, 3 runs, with standard deviation) of different methods on PathMNIST dataset with asymmetric noise}
 	\begin{tabular}{|c|c|c|c|c|}
 		\hline
 		Noise rate & 0.1 & 0.2 & 0.3 & 0.4 \\ \hline
 		baseline & 86.31±2.34 & 76.75±1.98 & 63.06±5.39 & 56.96±3.68 \\ \hline
 		O2U   & \textbf{90.68±1.02} & 89.36±1.06 & 86.5±0.62 & 72.0±0.9 \\ \hline
 		MixUp & 85.44±2.88 & 79.41±2.05 & 70.6±1.88 & 57.8±1.71 \\ \hline
 		coteaching+ & 89.68±1.35 & 88.62±1.48 & 83.19±0.51 & 79.38±2.86 \\ \hline
 		CDR   & 87.13±0.67 & 81.48±1.22 & 69.96±0.69 & 58.5±1.83 \\ \hline
 		Self-adaptive    & 88.84±0.96 & 81.46±0.51 & 71.24±2.18 & 57.55±1.56 \\ \hline
 		Multiclass & 85.76±0.61 & 78.66±1.29 & 69.01±0.51 & 57.27±0.59 \\ \hline
 		LNL\_SR & 85.25±2.51 & 86.91±1.13 & 87.49±0.43 & 86.42±1.12 \\ \hline
 		Ours  & 89.13±1.61 & \textbf{90.76±0.54} & \textbf{89.96±0.8} & \textbf{87.29±0.73} \\ \hline
 	\end{tabular}%
 	\label{accpatha}%
 \end{table}%

\indent Table \ref{accpaths} and tabel \ref{accpatha} shows the classification accuracy of different noise-robust deep learning methods on PathMNIST dataset with symmetric and asymmetric. Evidently, most of the existing noise-robust methods can more or less improve the accuracy score on this dataset, depending on the noise rate settings. For instance, Coteaching-Plus ranks first under symmetric noise when noise rate is 0.1, but fails to maintain this position in other conditions. Conversely, our original method win the first place across most of the noise rate settings, particularly when noise rate is very big. Nevertheless, when the noise rate is small, our method is still competitive and obtained sub-optimal performance.

%OCTMNIST acc table
\begin{table}[H]
	\centering
	\caption{Average accuracy (\%, 3 runs, with standard deviation) of different methods on OCTMNIST dataset with symmetric noise}
	\begin{tabular}{|c|c|c|c|c|c|}
		\hline
		Noise rate & 0 & 0.1 & 0.2 & 0.3 & 0.4 \\ \hline
		Baseline & 73.80±0.95 & 71.83±1.46 & 65.00±0.95 & 60.13±1.37 & 52.03±1.12 \\ \hline
		O2U   & 74.13±1.40 & 73.23±0.57 & 73.43±1.67 & 67.47±0.35 & 59.10±1.30 \\ \hline
		MixUp & 72.33±0.38 & 66.63±0.61 & 61.23±2.25 & 54.8±2.09 & 47.13±1.89 \\ \hline
		Coteaching+ & 75.97±0.67 & 74.23±2.08 & 69.20±0.66 & 66.33±1.29 & 62.10±2.71 \\ \hline
		CDR   & 75.50±1.13 & 70.57±2.23 & 65.00±0.95 & 59.93±1.60 & 47.30±1.73 \\ \hline
		Self-adaptive  & 74.37±0.40 & 71.60±0.66 & 66.23±0.23 & 61.37±0.40 & 51.37±1.44 \\ \hline
		Multiclass & 75.47±1.19 & 70.30±0.66 & 64.50±0.66 & 60.80±0.87 & 45.03±2.81 \\ \hline
		Lnl\_sr & 76.33±0.06 & 74.27±0.59 & 73.56±2.02 & 72.53±0.38 & 71.70±1.04 \\ \hline
		Ours  & \textbf{76.50±0.26} & \textbf{74.97±0.12} & \textbf{74.67±0.50} & \textbf{73.07±1.24} & \textbf{72.30±1.04} \\ \hline
	\end{tabular}%
	\label{accoct}%
\end{table}%

\indent Unlike PathMNIST, the classification results in table \ref{accoct} suggest that noisy OCTMNIST dataset is a much more difficult task for noise-robust learning methods. It can be observed from this table that compared with the non-robust baseline method, many existing noise-robust methods cannot really improve the classification accuracy on this dataset. For example, four out of the seven tested comparison methods failed to outperform the baseline method when the noise rate is 0.4. However, even for this difficult task, our method still obtained top performance, obtaining best accuracy score under all noise rate settings. This result proves the potential of our method in addressing the difficulties from more challenging noisy datasets.

%PneumoniaMNIST acc table
\begin{table}[H]
	\centering
	\caption{Average accuracy (\%, 3 runs, with standard deviation) of different methods on PneumoniaMNIST dataset with symmetric noise}
	\begin{tabular}{|c|c|c|c|c|c|}
		\hline
		Noise rate & 0 & 0.1 & 0.2 & 0.3 & 0.4 \\ \hline
		Baseline & 85.95±1.80 & 81.04±0.81 & 75.27±1.29 & 69.12±1.21 & 57.48±1.48 \\ \hline
		O2U   & 85.42±0.80 & 82.69±0.70 & 79.54±1.62 & 74.20±2.20 & 60.47±2.65 \\ \hline
		MixUp & \textbf{86.65±0.33} & 84.35±1.30 & 81.09±0.70 & 73.18±2.51 & 61.91±4.69 \\ \hline
		Coteaching+ & 85.10±1.95 & 81.36±0.82 & 81.36±2.60 & 76.65±1.64 & 71.10±10.7 \\ \hline
		CDR   & 85.15±0.09 & 82.00±0.18 & 77.25±3.16 & 68.53±1.29 & 60.31±2.59 \\ \hline
		Self-adaptive  & 85.79±0.96 & 79.76±1.57 & 73.77±2.98 & 65.92±0.24 & 57.75±2.40 \\ \hline
		Multiclass & 84.29±0.43 & 81.20±1.22 & 76.92±1.21 & 71.10±1.69 & 60.58±3.24 \\ \hline
		Lnl\_sr & 85.37±0.18 & 84.24±0.25 & 85.1±0.55 & 81.03±2.98 & 65.22±1.16 \\ \hline
		Ours  & 84.83±0.61 & \textbf{85.04±0.89} & \textbf{85.15±1.49} & \textbf{83.65±2.12} & \textbf{75.53±0.56} \\ \hline
	\end{tabular}%
	\label{accpneumonia}%
\end{table}%

\indent The experimental results of PneumoniaMNIST are reported in table \ref{accpneumonia}. For this binary-classification task, The performance of most comparison methods showed a sharp decrease when noise rate is very big. For example, when noise rate rises from 0.3 to 0.4, a rapid decline in accuracy score can be observed in all existing methods, such as the none-robust baseline method (11.7\%) LNL\_SR (15.8\%).
\\
\indent Our proposed method maintained excellent accuracy score even when noise rate is very big. It is noticeable that as long as label noise existed (noise rate from 0.1 to 0.4), our method ranked first in all the experiments. Meanwhile, when the dataset was clean, MixUp obtained the highest accuracy score, indicating the efficacy of sample augmentation on clean datasets.

\subsection{Experiments on 3D benchmark datasets}

\indent In this section, we present the experimental results on  three-dimensional medical image datasets, including OrganMNIST3D and VesselMNIST3D. We introduced symmetric noise to both datasets and also implemented asymmetric noise on OrganMNIST3D. From table \ref{accorgans} to \ref{accvessel}, it is observable that the test accuracy of 3D medical images show a similar decrease on noisy labels to what happens on 2D medical images.

%acc table organ symmetric
\begin{table}[H]
	\centering
	\caption{Average accuracy (\%, 3 runs, with standard deviation) of different methods on OrganMNIST dataset with symmetric noise}
	\begin{tabular}{|c|c|c|c|c|c|}
		\hline
		Noise rate & 0 & 0.1 & 0.2 & 0.3 & 0.4 \\ \hline
		Baseline & 90.49±0.43 & 83.99±0.81 & 79.40±1.92 & 68.96±1.58 & 59.45±3.09 \\ \hline
		O2U   & 82.19±0.90 & 84.75±0.72 & 85.30±1.15 & 74.21±2.50 & 66.83±2.30 \\ \hline
		MixUp & \textbf{92.51±1.51} & 84.44±0.68 & 77.83±1.42 & 68.58±3.78 & 62.19±2.61 \\ \hline
		Coteaching+ & 86.50±5.11 & 85.96±0.78 & 84.59±2.32 & 77.38±2.01 & 66.01±2.46 \\ \hline
		CDR   & 91.48±0.85 & 81.64±1.70 & 74.97±2.08 & 68.09±1.23 & 58.20±2.27 \\ \hline
		Self-adaptive  & 87.92±1.47 & 79.84±0.16 & 69.45±0.53 & 64.21±1.07 & 54.10±1.99 \\ \hline
		Multiclass & 89.89±1.62 & 83.28±1.28 & 76.83±1.81 & 71.42±4.76 & 58.90±0.81 \\ \hline
		Lnl\_sr & 89.56±0.38 & 86.34±1.39 & 82.68±1.81 & 79.13±4.07 & 68.98±2.59 \\ \hline
		Ours  & 89.40±1.27 & \textbf{87.16±0.41} & \textbf{85.79±0.09} & \textbf{81.37±1.48} & \textbf{76.06±1.66} \\ \hline
	\end{tabular}%
	\label{accorgans}%
\end{table}%

% Acc table Organ asymmetric noise
\begin{table}[H]
	\centering
	\caption{Average accuracy (\%, 3 runs, with standard deviation) of different methods on OrganMNIST dataset with asymmetric noise}
	\begin{tabular}{|c|c|c|c|c|}
		\hline
		Noise rate & 0.1 & 0.2  & 0.3 & 0.4 \\ \hline
		baseline & 85.57±2.61 & 75.9±1.71 & 69.02±3.53 & 53.77±2.9 \\ \hline
		O2U   & 84.05±1.52 & 81.64±1.57 & 72.02±3.05 & 61.15±1.3 \\ \hline
		mixup & 85.08±1.28 & 76.29±1.49 & 65.52±1.24 & 54.15±2.0 \\ \hline
		coteaching & 86.12±1.09 & 77.82±3.75 & 67.38±5.15 & 55.74±4.89 \\ \hline
		cdr   & 82.46±1.0 & 73.22±2.79 & 66.01±1.55 & 56.5±1.84 \\ \hline
		self  & 79.56±1.87 & 70.44±3.14 & 63.61±2.64 & 55.57±2.8 \\ \hline
		multiclass & 84.05±0.76 & 75.25±1.07 & 67.1±2.11 & 55.9±0.43 \\ \hline
		lnlsr & 86.23±0.43 & 79.02±1.77 & 67.05±0.71 & 58.58±2.58 \\ \hline
		ours  & \textbf{86.45±1.27} & \textbf{83.5±1.94} & \textbf{74.65±1.92} & \textbf{64.26±2.05} \\ \hline
	\end{tabular}%
	\label{accorgana}%
\end{table}%

\indent Table \ref{accorgans} and table \ref{accorgana} presents the outcomes of original method and other competitors on OrganMNIST3D dataset under symmetric and asymmetric noise. In this challenging 3D abdominal CT classification task, our method obtained sub-optimal results on a clean dataset, and as noise rate increases, the advantage of our method is even more apparent, outperforming its competitors across all noise rate settings. It also can be observed from table \ref{accorgana} that our original method still ranked first on the asymmetric noise conditions.

%acc table Vessel
\begin{table}[H]
	\centering
	\caption{Average accuracy (\%, 3 runs, with standard deviation) of different methods on VesselMNIST dataset with symmetric noise}
	\begin{tabular}{|c|c|c|c|c|c|}
		\hline
		Noise rate & 0 & 0.1 & 0.2 & 0.3 & 0.4 \\ \hline
		Baseline & 90.84±1.36 & 87.00±2.03 & 82.11±1.49 & 68.76±1.68 & 62.83±4.65 \\ \hline
		O2U   & 90.75±1.53 & 90.49±1.06 & 79.23±1.09 & 70.33±3.57 & 58.55±1.75 \\ \hline
		MixUp & 92.15±1.36 & 84.91±1.58 & 77.49±5.50 & 68.06±1.89 & 59.42±4.21 \\ \hline
		Coteaching+ & 88.48±0.26 & \textbf{91.01±1.34} & 85.69±3.54 & 76.97±2.24 & 64.83±2.11 \\ \hline
		CDR   & 92.32±0.15 & 87.26±0.76 & 80.28±2.88 & 71.29±1.66 & 62.22±1.66 \\ \hline
		Self-adaptive  & 91.62±1.20 & 83.94±1.06 & 79.15±2.42 & 73.04±11.6 & 56.89±3.93 \\ \hline
		Multiclass & \textbf{92.67±1.05} & 89.79±0.26 & 81.85±2.63 & 72.34±3.71 & 64.40±1.63 \\ \hline
		LNL\_SR & 91.27±0.54 & 89.88±0.40 & 87.00±1.06 & 83.34±5.26 & 73.12±1.29 \\ \hline
		Ours  & 89.70±0.65 & 90.23±0.84 & \textbf{88.48±1.45} & \textbf{85.95±1.44} & \textbf{83.24±2.77} \\ \hline
	\end{tabular}%
	\label{accvessel}%
\end{table}%

\indent VesselMNIST is another imbalance 3D dataset, which is also a challenge for all methods. We can see that our method successfully obtained the higher score on this difficult dataset.

\indent From table \ref{accvessel}, it can be observed that in VesselMNSIT dataset, our method still ranked top in all the experiments. Especially, when the noise rate is big, our method surpassed the second-best method by more than 10\% in classification accuracy. This robust performance demonstrates our method’s ability of finding a good way to learn the correct knowledge from heavily noisy samples. Furthermore, even in situations with small label noise, our method still obtained an acceptable classification performance.

\indent This study marks the first exploration of whether current noise-robust deep learning methods still work on 3D medical images. The results in this experiment reveal that the efficacy of noise-robust learning methods is not solely determined by the dimensionality, since most of the existing noise-robust learning methods designed for 2D images still exhibit varying degrees of efficacy on 3D medical images. Among them, our original method again obtains the highest level of robustness, particularly in handling extremely noisy data.  This study substantiates the potential effectiveness of noise-robust learning in the context of 3D medical images, which makes up a big proportion of contemporary medical applications.

\subsection{Experiments on a real-world noisy dataset}

\indent In this section, we introduce our experiment results on the NoisyCXR dataset. Unlike above benchmard datasets where we could implement different types ratios of artificial noise to produce results under all settings, for this dataset there are only two situations: clean (noise rate = 0) and noisy (noise rate is around 0.4). 

% ACC noisycxr
\begin{table}[H]
	\centering
	\caption{Average accuracy (\%, 3 runs, with standard deviation) of different methods on NoisyCXR dataset}
	\begin{tabular}{|c|c|c|}
		\hline
		Noise & Clean & Noisy \\ \hline
		Baseline  & 79.95±0.33 & 59.15±0.44 \\ \hline
		O2U   & 81.65±0.42 & 61.76±0.88 \\ \hline
		MixUp & 80.17±0.31 & 61.98±0.71 \\ \hline
		Coteaching-plus    & 81.84±0.14 & 62.34±1.14 \\ \hline
		CDR   & 80.74±1.06 & 61.39±0.54 \\ \hline
		Self-adaptive  & 80.22±0.53 & 57.68±0.65 \\ \hline
		Multiclass & 80.19±0.32 & 55.54±0.36 \\ \hline
		LNL\_SR & 80.11±0.17 & 60.49±3.4 \\ \hline
		ours  & \textbf{81.94±0.12} & \textbf{66.9±0.46} \\ \hline
	\end{tabular}%
	\label{accnoisycxr}%
\end{table}%

\indent Nevertheless, table \ref{accnoisycxr} reveals that the presence of label noise in the training set significantly impairs the generalization performance on the test set for all tested methods. Notably, the original method ranked first a significant lead in noisy labels, and also obtained the first place with a small lead on the clean set. These results again confirmed that our method is beyond the scope of MNIST-like simplistic benchmark datasets, and underscores the potential applicability of our method in real-world medical image application.

%4.9 Performance of noise rate estimation module, Ablation study
\subsection{Evaluation and ablation study of noise rate estimation module}

\indent In this section, we conducted a comparison between the predicted noise rate generated by our module and the actual noise rate, aiming to assess the prediction ability of this module. Ablation study will also be carried out in this section to validation the performance of our model with and without the linear regression noise rate estimation module.

\indent The detailed noise-rate prediction results are presented in table \ref{noiseratepredictionresults}, showing that for most datasets, our predictions closely align with the actual noise rate. Notably, for the OrganMNIST dataset, when noise rate 0, our model exhibit a mean prediction of 0.165, which is quite different from the actual noise rate 0. The reason for the lesser precise predictions on organMNIST may be attributed to its having the highest number of categories (11 classes), which the training of linear regression model has not adequately covered.

\indent Despite OrganMNIST, the difference is still in an accepted range. These findings prove the effectiveness of our noise rate estimation module. Regarding the NoisyCXR dataset, although the predicted average noise rate under 'clean' (RSNA) labels is 0.114, which seems to have a certain discrepancy from 0, it is important to note that RSNA labels are not necessarily completely accurate and might contain a minimal amount of incorrect labels. In this dataset, it is visibly apparent that some pulmonary inflammatory lesions have not been classified as pneumonia-like opacity by RSNA, and hence, were not categorized as pneumonia. Furthermore, our model achieved the highest score under the assumption of being clean, whereas the baseline model, which did not employ any noise-robust methods, performed the worst. This result also suggests that RSNA labels may contain a minimal amount of label noise.

%Noise rate prediction results
\begin{table}[H]
	\centering
	\caption{Predicted noise rate on different datasets (mean of three runs)}
	\begin{tabular}{|c|c|c|c|c|c|}
		\hline
		Actual noise rate	&0 			&0.1  		&0.2 		&0.3   		&0.4		\\ \hline
		Path (Symmetric)	&0.005 		&0.079 		&0.173 		&0.267 		&0.362		\\ \hline
		Path (Asymmetric)	&/ 			&0.078 		&0.147 		&0.230 		&0.234		\\ \hline
		OCT (Symmetric)		&0 			&0.133 		&0.226 		&0.340 		&0.449		\\ \hline
		Pneumonia 			&0 			&0.068 		&0.162 		&0.257 		&0.382		\\ \hline
		Organ (Symmetric)	&0.165 		&0.176 		&0.256 		&0.316 		&0.364		\\ \hline
		Organ (Asymmetric)	&/ 			&0.199 		&0.246 		&0.367 		&0.415		\\ \hline
		Vessel 				&0.072 		&0.142 		&0.236 		&0.265 		&0.392		\\ \hline
		NoisyCXR 			&0.114 		&/ 			&/ 			&/ 			&0.379 		\\ \hline
	\end{tabular}
	\label{noiseratepredictionresults}
\end{table}

\indent It can be seen from the above results that our noise-rate estimation module can offer relatively accurate noise rate prediction across various types of noise types and noise rates. Currently, this linear regressor is trained on five MNIST like small datasets. Moving forward, if the training data is sufficient, we could consider employing other machine learning models, such as an extra neural network to predict the noise rate.
\\
\indent However, this also highlights a limitation of our study: we need to artificially control the noise rate in other auxiliary datasets first, allowing our model to learn the loss distribution under different noise types and rates. In other words, this noise rate prediction module requires some auxiliary datasets for preliminary knowledge. Nonetheless, in the absence of other auxiliary datasets, we could still consider omitting the noise rate estimation module and solely use other modules with a fixed forget rate. The following section conducts an ablation study of how much does the noise-rate estimator help to improve the classification accuracy, compared with simply using a fixed forget rate.

%Ablation study between constant and estimated forget rate
\begin{table}[H]
	\centering
	\caption{Comparison of overall accuracy score of between constant and estimated forget rate on MedMNIST datasets}
	\begin{tabular}{|c|c|c|c|c|c|c|}
		\hline
		Forget rate & 0 & 0.1 & 0.2 & 0.3 & 0.4 & Estimated \\ \hline
		Path  & 88.22 & 88.12 & 88.06 & 89.2  & 89.27 & 88.58 \\ \hline
		OCT   & 72.74 & 72.02 & 73.44 & 72.96 & 73.56 & 74.3 \\ \hline
		Pneumonia & 79.01 & 80.32 & 81.54 & 80.42 & 77.76 & 82.97 \\ \hline
		Organ & 81.61 & 82.26 & 81.48 & 83.9  & 83.15 & 83.96 \\ \hline
		Vessel & 85.34 & 82.67 & 87.22 & 88.74 & 86.75 & 87.52 \\ \hline
	\end{tabular}%
	\label{ablationstudy}%
\end{table}%

%\subsection{Ablation study}
\indent Table \ref{ablationstudy} presents the results of the ablation study with different forget rates across the five MedMNIST datasets. We set five different fixed forget rates (0 to 0.4) and compared the overall classification accuracy of our original noise-robust learning method of the fixed forget rates and the estimated one. 
\\
\indent The findings indicate an improvement in overall classification accuracy dues to our adaptive noise rate estimation module when compared to a fixed noise rate. Notably, with the integration of this noise rate estimation module, the overall accuracy outperforms that of small forget rates (0 and 0.1) on all datasets, and outperforms big forget rates (0.2, 0.3, 0.4) on most datasets. These results signify the potential of this noise-rate estimation module to enhance practical applicability of our sample selection method, showing potential in clinical datasets where the noise rate of a dataset is unknown.

%5. Conclusion
\section{Conclusion}
\indent In this paper, we present a noise-robust deep learning method designed for the classification of diverse medical images with label noise. This method consists of three modules tailored for noise robustness: noise rate estimation, sample selection, and sparse regularization. Experiments with both 2D and 3D medical image datasets were conducted to evaluate the proposed method. The outcomes demonstrate the efficacy of this method in enhancing classification performance across varying levels of label noise, particularly in severely noisy datasets.

%References
\bibliographystyle{plain}
\bibliography{mybib}

\appendix

\section{F1 scores of all methods}

\indent Figure \ref{f1paths} to figure \ref{f1vessel} are the F1 score of all tested methods on the test set. For MedMNIST datasets, results come from noise rate 0.2, and for NoisyCXR, noise rate is around 0.4. It can be observed that the original method kept the lead in terms of F1 score, which corresponds with the results from the accuracy score.

% Path Symmetric
\begin{table}[H]
	\centering
	\caption{F1 scores of all tested methods on PathMNIST dataset under symmetric noise rate 0.2}
	\begin{tabular}{|c|c|c|c|c|c|}
		\hline
		Noise rate & 0     & 0.1   & 0.2   & 0.3   & 0.4 \\ \hline
		Baseline & 88.45±2.32 & 85.77±0.55 & 78.56±1.42 & 70.59±0.2 & 63.9±1.73 \\ \hline
		O2U   & 90.8±0.41 & 90.28±0.67 & 88.99±0.88 & 86.91±0.4 & 80.57±0.83 \\ \hline
		MixUp & 87.7±0.78 & 85.62±0.89 & 83.13±1.49 & 77.5±1.82 & 68.51±1.03 \\ \hline
		Coteacing+ & 89.24±1.32 & 90.2±0.26 & 88.71±0.19 & 84.54±1.11 & 84.3±0.68 \\ \hline
		CDR   & 90.29±0.32 & 86.92±0.87 & 83.62±0.07 & 77.16±2.45 & 67.0±0.71 \\ \hline
		Self-adaptive & 89.99±1.56 & 87.59±0.98 & 83.89±0.52 & 78.74±0.68 & 69.19±1.18 \\ \hline
		Multiclass & 89.32±1.13 & 85.12±1.11 & 80.61±0.51 & 71.84±0.77 & 62.21±0.41 \\ \hline
		LNL\_SR & 87.98±1.49 & 87.8±1.42 & 86.07±0.95 & 86.88±0.58 & 86.33±1.03 \\ \hline
		ours  & 89.95±1.38 & 88.24±0.88 & 88.2±1.83 & 88.68±0.86 & 87.42±0.46 \\ \hline
	\end{tabular}%
	\label{f1paths}%
\end{table}%

% Path ASymmetric
\begin{table}[H]
	\centering
	\caption{F1 scores of all tested methods on PathMNIST dataset under asymmetric noise rate 0.2}
	\begin{tabular}{|c|c|c|c|c|}
		\hline
		Noise rate &  0.1   & 0.2   & 0.3   & 0.4 \\ \hline
		baseline & 86.07±2.35 & 76.8±2.31 & 63.45±5.28 & 58.01±4.34 \\ \hline
		O2U   & 90.22±1.01 & 89.06±1.08 & 86.32±0.58 & 72.94±0.84 \\ \hline
		mixup & 85.5±2.63 & 79.59±2.06 & 71.22±1.7 & 59.02±1.72 \\ \hline
		coteaching & 89.48±1.34 & 88.69±1.53 & 83.32±0.5 & 79.92±2.97 \\ \hline
		cdr   & 86.92±0.63 & 81.38±1.12 & 70.64±0.5 & 59.88±1.99 \\ \hline
		self  & 88.63±0.87 & 81.65±0.54 & 72.01±1.98 & 59.13±1.79 \\ \hline
		multiclass & 85.53±0.61 & 78.69±1.23 & 69.82±0.55 & 59.04±0.59 \\ \hline
		lnlsr & 85.36±2.23 & 86.95±1.12 & 87.43±0.29 & 86.49±1.01 \\ \hline
		ours  & 89.08±1.56 & 90.66±0.54 & 89.87±0.72 & 87.5±0.67 \\ \hline
	\end{tabular}%
	\label{f1patha}%
\end{table}%

% OCT symmetric'
\begin{table}[H]
	\centering
	\caption{F1 scores of all tested methods on OCTMNIST dataset under symmetric noise rate 0.2}
	\begin{tabular}{|c|c|c|c|c|c|}
		\hline
		Noise rate & 0     & 0.1   & 0.2   & 0.3   & 0.4 \\ \hline
		Baseline & 70.63±1.94 & 68.51±3.51 & 62.06±2.78 & 57.78±3.94 & 50.69±1.04 \\ \hline
		O2U   & 69.47±0.61 & 69.22±0.83 & 70.32±2.4 & 64.04±0.63 & 56.3±1.93 \\ \hline
		MixUp & 67.21±0.8 & 61.29±0.72 & 57.11±2.91 & 50.75±2.86 & 44.15±1.91 \\ \hline
		Coteacing+ & 73.51±0.63 & 71.4±3.09 & 65.92±0.74 & 63.79±1.87 & 59.86±2.46 \\ \hline
		CDR   & 72.54±1.39 & 67.0±2.81 & 61.56±1.0 & 57.11±1.87 & 45.03±2.13 \\ \hline
		Self-adaptive & 71.04±0.55 & 68.63±0.69 & 63.16±0.29 & 59.2±0.18 & 50.22±1.15 \\ \hline
		Multiclass & 72.32±1.34 & 67.61±0.83 & 62.37±1.21 & 59.14±0.8 & 44.09±2.31 \\ \hline
		LNL\_SR & 73.75±0.28 & 71.33±0.66 & 70.99±2.42 & 69.8±0.86 & 69.16±1.21 \\ \hline
		ours  & 73.93±0.29 & 72.5±0.14 & 72.09±0.61 & 70.54±1.44 & 69.28±1.53 \\ \hline
	\end{tabular}%
	\label{f1oct}%
\end{table}%

% Pneumonia symmetric
\begin{table}[H]
	\centering
	\caption{F1 scores of all tested methods on PneumoniaMNIST dataset under symmetric noise rate 0.2}
	\begin{tabular}{|c|c|c|c|c|c|}
		\hline
		Noise rate & 0     & 0.1   & 0.2   & 0.3   & 0.4 \\ \hline
		Baseline & 85.04±2.08 & 79.81±0.79 & 73.54±1.84 & 69.01±1.37 & 57.9±1.39 \\ \hline
		O2U   & 84.45±0.92 & 81.48±0.9 & 78.39±2.12 & 73.9±2.49 & 60.91±2.57 \\ \hline
		MixUp & 85.97±0.36 & 83.63±1.32 & 80.55±0.79 & 72.97±2.14 & 62.29±4.69 \\ \hline
		Coteacing+ & 84.21±2.2 & 85.17±1.08 & 80.23±2.89 & 75.42±1.43 & 70.77±9.82 \\ \hline
		CDR   & 84.57±0.62 & 80.55±0.44 & 77.56±1.66 & 67.71±2.27 & 61.77±2.23 \\ \hline
		Self-adaptive & 84.99±1.06 & 78.64±1.63 & 73.06±3.11 & 65.78±0.55 & 58.03±2.52 \\ \hline
		Multiclass & 83.27±0.48 & 79.93±1.58 & 75.78±1.96 & 70.54±2.31 & 61.12±3.11 \\ \hline
		LNL\_SR & 84.42±0.22 & 83.21±0.25 & 84.25±0.6 & 79.85±3.14 & 65.04±1.04 \\ \hline
		ours  & 83.8±0.69 & 85.09±1.01 & 85.04±1.82 & 82.7±2.41 & 73.14±1.68 \\ \hline
	\end{tabular}%
	\label{f1pneumonia}%
\end{table}%

% Organ Symmetric
\begin{table}[H]
	\centering
	\caption{F1 scores of all tested methods on OrganMNIST dataset under symmetric noise rate 0.2}
	\begin{tabular}{|c|c|c|c|c|c|}
		\hline
		Noise rate & 0     & 0.1   & 0.2   & 0.3   & 0.4 \\ \hline
		Baseline & 90.49±0.46 & 83.93±0.92 & 79.28±1.9 & 68.65±1.78 & 59.33±3.08 \\ \hline
		O2U   & 82.15±0.89 & 84.68±0.7 & 85.3±1.18 & 74.04±2.43 & 66.72±2.5 \\ \hline
		MixUp & 92.46±1.52 & 84.25±0.7 & 77.4±1.35 & 68.17±3.7 & 62.12±2.53 \\ \hline
		Coteacing+ & 85.03±7.73 & 86.15±0.39 & 84.61±2.23 & 77.33±2.04 & 65.87±2.42 \\ \hline
		CDR   & 90.75±1.11 & 82.15±2.29 & 75.67±1.91 & 68.9±1.41 & 57.76±1.85 \\ \hline
		Self-adaptive & 87.83±1.47 & 79.76±0.16 & 69.24±0.5 & 64.21±0.96 & 54.31±2.38 \\ \hline
		Multiclass & 89.91±1.6 & 83.09±1.27 & 76.66±1.76 & 71.2±4.6 & 58.93±0.54 \\ \hline
		LNL\_SR & 89.57±0.35 & 86.31±1.46 & 82.48±1.72 & 78.86±4.11 & 68.73±2.36 \\ \hline
		ours  & 89.37±1.22 & 87.04±0.48 & 85.28±11.44 & 81.3±1.59 & 76.06±1.69 \\ \hline
	\end{tabular}%
	\label{f1organs}%
\end{table}%

% Organ Asymmetric
\begin{table}[H]
	\centering
	\caption{F1 scores of all tested methods on OrganMNIST dataset under asymmetric noise}
	\begin{tabular}{|c|c|c|c|c|}
		\hline
		Noise rate &  0.1 & 0.2 &0.3 & 0.4 \\ \hline
		baseline & 85.57±2.58 & 76.06±1.69 & 69.35±3.55 & 54.17±3.11 \\ \hline
		O2U   & 84.25±1.42 & 81.57±1.41 & 71.94±3.22 & 61.54±1.04 \\ \hline
		mixup & 84.98±1.26 & 76.26±1.47 & 65.7±1.35 & 54.6±2.01 \\ \hline
		coteaching & 85.71±1.77 & 76.51±5.61 & 67.16±6.0 & 53.96±5.53 \\ \hline
		cdr   & 82.45±0.93 & 73.29±2.84 & 66.22±1.69 & 56.82±2.32 \\ \hline
		self  & 79.58±1.8 & 70.62±3.14 & 64.05±2.48 & 56.34±2.87 \\ \hline
		multiclass & 84.02±0.75 & 75.4±0.92 & 67.17±2.15 & 56.46±0.43 \\ \hline
		lnlsr & 86.23±0.43 & 79.02±1.77 & 67.05±0.71 & 58.58±2.58 \\ \hline
		ours  & 86.39±1.35 & 83.39±1.9 & 74.35±1.79 & 64.2±1.94 \\ \hline
	\end{tabular}%
	\label{f1organa}%
\end{table}%

% Vessel Symmetric
\begin{table}[H]
	\centering
	\caption{F1 scores of all tested methods on VesselMNIST dataset under symmetric noise}
	\begin{tabular}{|c|c|c|c|c|c|}
		\hline
		Noise rate & 0     & 0.1   & 0.2   & 0.3   & 0.4 \\ \hline
		Baseline & 89.21±2.91 & 86.76±1.7 & 83.1±1.09 & 73.81±1.24 & 69.63±3.74 \\ \hline
		O2U   & 87.73±3.19 & 87.85±1.88 & 81.52±0.62 & 75.48±2.68 & 66.14±1.47 \\ \hline
		MixUp & 91.66±1.39 & 86.12±1.18 & 80.39±3.6 & 73.47±1.4 & 66.77±3.33 \\ \hline
		Coteacing+ & 83.62±0.54 & 89.36±1.89 & 84.78±2.76 & 79.33±1.58 & 71.15±1.68 \\ \hline
		CDR   & 91.83±0.07 & 87.06±0.9 & 81.24±1.63 & 76.22±1.46 & 69.85±1.29 \\ \hline
		Self-adaptive & 91.02±1.53 & 85.65±0.75 & 81.32±1.76 & 75.98±6.44 & 64.71±3.36 \\ \hline
		Multiclass & 91.61±1.28 & 89.07±0.5 & 83.55±2.09 & 76.95±2.73 & 70.75±1.22 \\ \hline
		LNL\_SR & 89.41±0.68 & 88.79±0.56 & 86.22±1.03 & 84.08±3.28 & 77.09±1.11 \\ \hline
		ours  & 89.04±1.19 & 89.91±1.6 & 87.11±0.75 & 85.04±1.28 & 82.72±2.31 \\ \hline
	\end{tabular}%
	\label{f1vessel}%
\end{table}%

% NoisyCXR
\begin{table}[H]
	\centering
	\caption{F1 scores of all tested methods on NoisyCXR dataset}
	\begin{tabular}{|c|c|c|}
		\hline
		Noise & Clean & Noisy \\ \hline
		base  & 79.36±0.3 & 62.45±0.53 \\ \hline
		o2u   & 80.47±0.4 & 64.84±0.86 \\ \hline
		mixup & 78.75±0.24 & 65.03±0.73 \\ \hline
		co    & 80.55±0.24 & 65.38±1.18 \\ \hline
		cdr   & 80.02±0.94 & 64.64±0.57 \\ \hline
		self  & 79.14±0.91 & 57.61±0.71 \\ \hline
		multi & 79.84±0.42 & 58.89±0.26 \\ \hline
		lnlsr & 79.01±0.19 & 62.73±1.58 \\ \hline
		ours  & 80.80±0.33 & 69.53±0.48 \\ \hline
	\end{tabular}%
	\label{tab:addlabel}%
\end{table}%

\section{Training curves of all methods under noise rate 0.2}

\begin{figure}[H]
	\centering
	\includegraphics[scale=0.42]{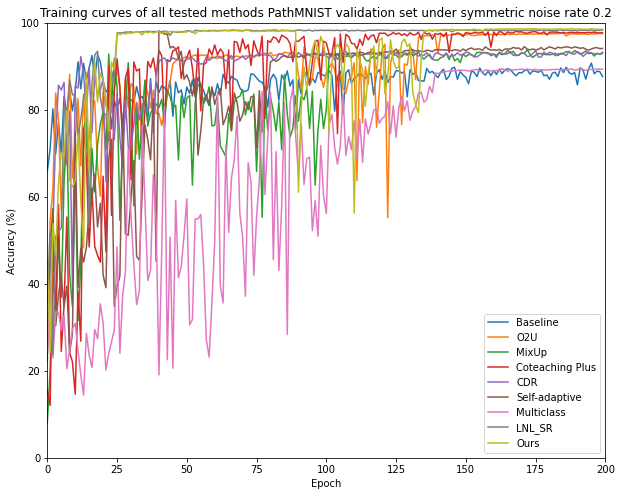}
	\caption{Training curves on PathMNIST with symmetric noise}
	\label{trainingpaths}
\end{figure}

\begin{figure}[H]
	\centering
	\includegraphics[scale=0.42]{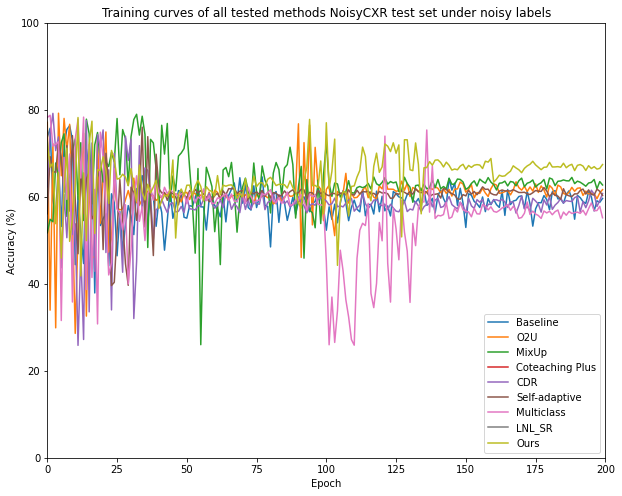}
	\caption{Training curves on NoisyCXR}
	\label{trainingnoisycxr}
\end{figure}

\indent Figure \ref{trainingpaths} to figure \ref{trainingnoisycxr} are the training curves of all tested methods on the validation set under noise rate 0.2. It can be observed that all methods have converged when training stops at epoch 200.

%training curve path symmetric

%training curve path asymmetric
\begin{figure}[H]
	\centering
	\includegraphics[scale=0.42]{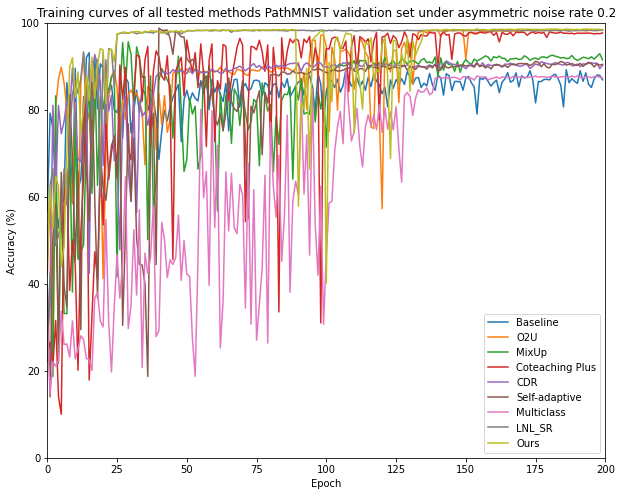}
	\caption{Training curves on PathMNIST with asymmetric noise}
	\label{trainingpatha}
\end{figure}

\begin{figure}[H]
	\centering
	\includegraphics[scale=0.42]{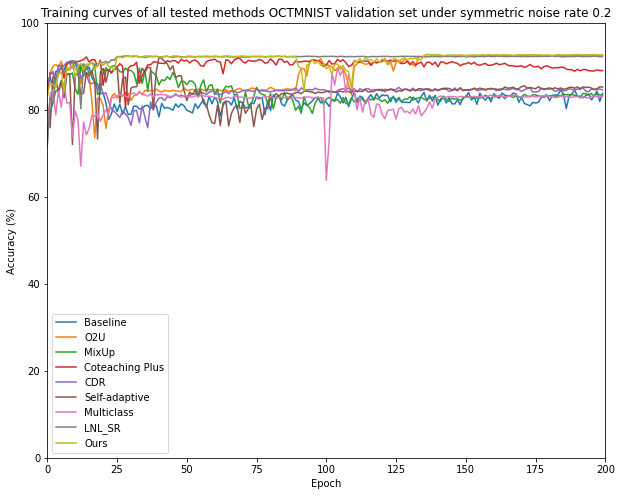}
	\caption{Training curves on OCTMNIST}
	\label{tcoct}
\end{figure}

\begin{figure}[H]
	\centering
	\includegraphics[scale=0.42]{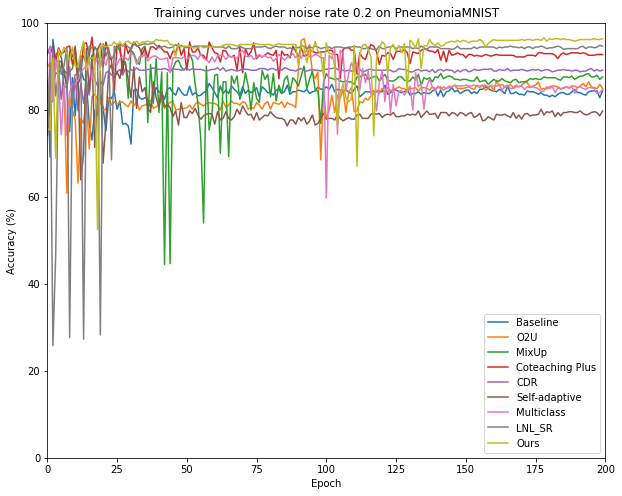}
	\caption{Training curves on PneumoniaMNIST}
	\label{tcpneumonia}
\end{figure}

\begin{figure}[H]
	\centering
	\includegraphics[scale=0.42]{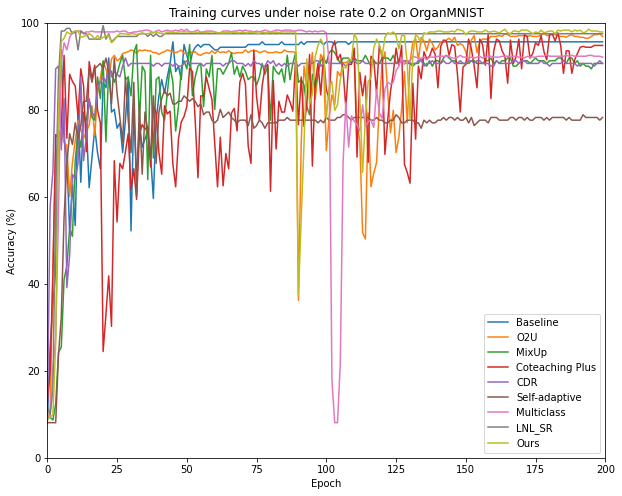}
	\caption{Training curves on OrganMNIST with symmetric noise}
	\label{trainingorgans}
\end{figure}

\begin{figure}[H]
	\centering
	\includegraphics[scale=0.42]{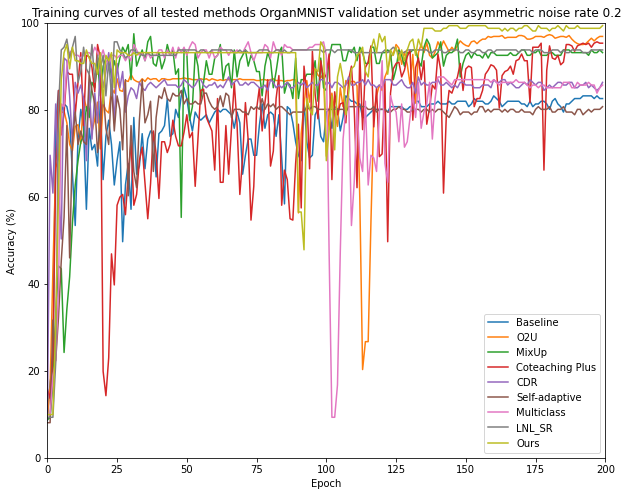}
	\caption{Training curves on OrganMNIST with asymmetric noise}
	\label{trainingorgana}
\end{figure}

\begin{figure}[H]
	\centering
	\includegraphics[scale=0.42]{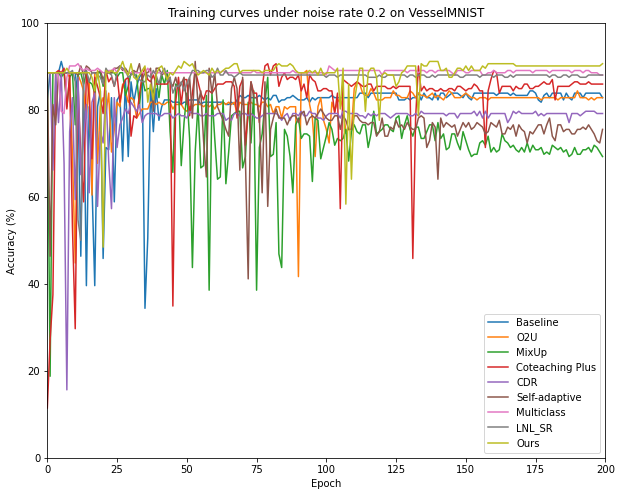}
	\caption{Training curves on VesselMNIST}
	\label{trainingvessel}
\end{figure}

%这个不用改了，training curve是0.4上的，0.0那个在改，这个不用改的。

\section{ROC curves of all tested methods at noise rate 0.2}

\begin{figure}[H]
	\centering
	\includegraphics[scale=0.42]{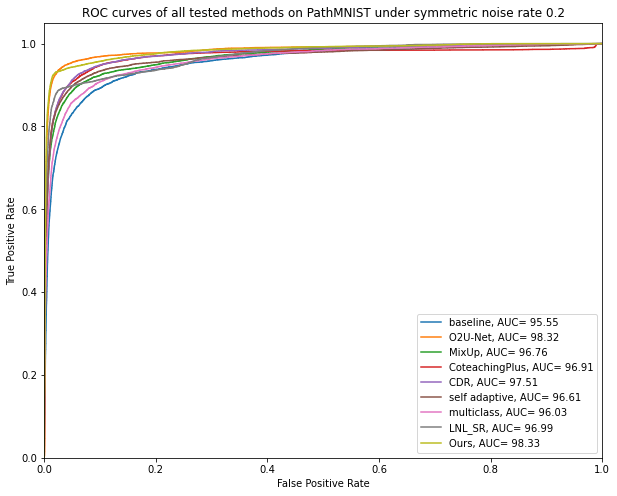}
	\caption{ROC curve and AUC of different methods on PathMNIST dataset under symmetric noise rate 0.2}
	\label{rocpaths}
\end{figure}

\begin{figure}[H]
	\centering
	\includegraphics[scale=0.42]{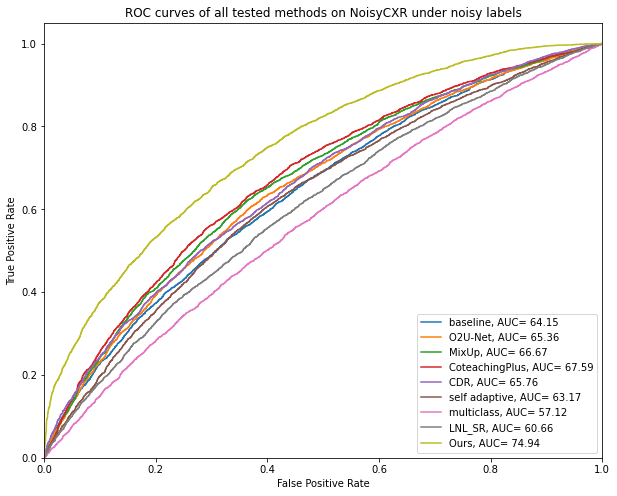}
	\caption{ROC curve and AUC of different methods on NoisyCXR test dataset under noise labels}
	\label{rocnoisycxr}
\end{figure}

\indent Figure \ref{rocpaths} to figure \ref{rocnoisycxr} are the ROC curves of all tested methods on the test set. For MedMNIST dataset, results are all from noise rate 0.2, while for NoisyCXR, noise rate is around 0.4. It can be observed that the original method kept the lead in terms of ROC curve and AUC score, which corresponds with the results from the accuracy score.

%ROC Path symmetric

%ROC Path asymmetric
\begin{figure}[H]
	\centering
	\includegraphics[scale=0.42]{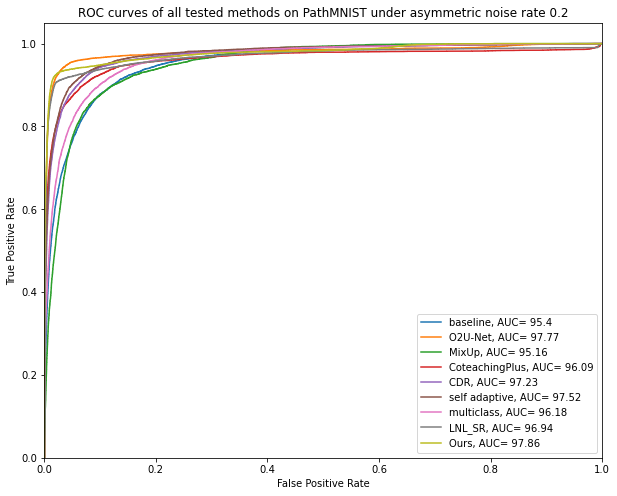}
	\caption{ROC curve and AUC of different methods on PathMNIST dataset under asymmetric noise rate 0.2}
	\label{rocpatha}
\end{figure}

%ROC oct
\begin{figure}[H]
	\centering
	\includegraphics[scale=0.42]{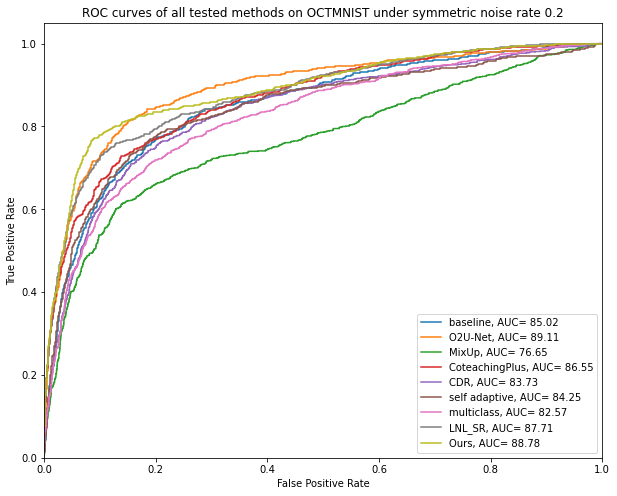}
	\caption{ROC curve and AUC of different methods on OCTMNIST dataset under symmetric noise rate 0.2}
	\label{rococts}
\end{figure}

%ROC pneumonia
\begin{figure}[H]
	\centering
	\includegraphics[scale=0.42]{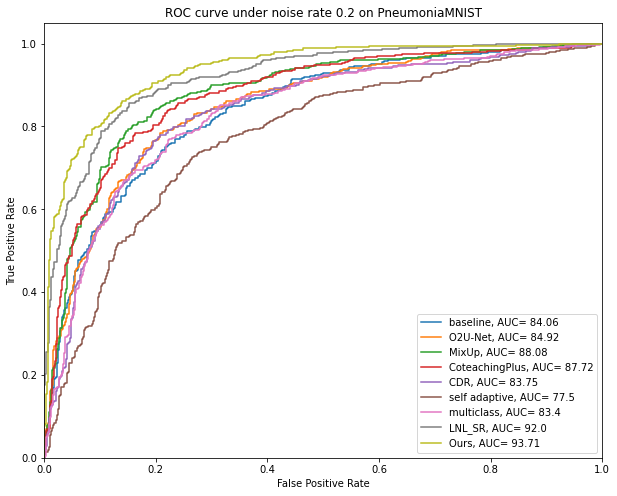}
	\caption{ROC curve and AUC of different methods on PneumoniaMNIST dataset under noise rate 0.2}
	\label{rocpneumonias}
\end{figure}

%roc organ symmetric
\begin{figure}[H]
	\centering
	\includegraphics[scale=0.42]{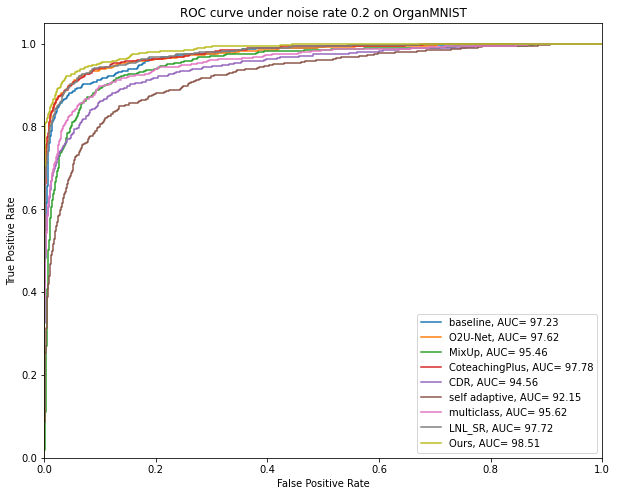}
	\caption{ROC curve and AUC of different methods on OrganMNIST3D dataset under symmetric noise rate 0.2}
	\label{rocorgans}
\end{figure}

%roc organ asymmetric
\begin{figure}[H]
	\centering
	\includegraphics[scale=0.42]{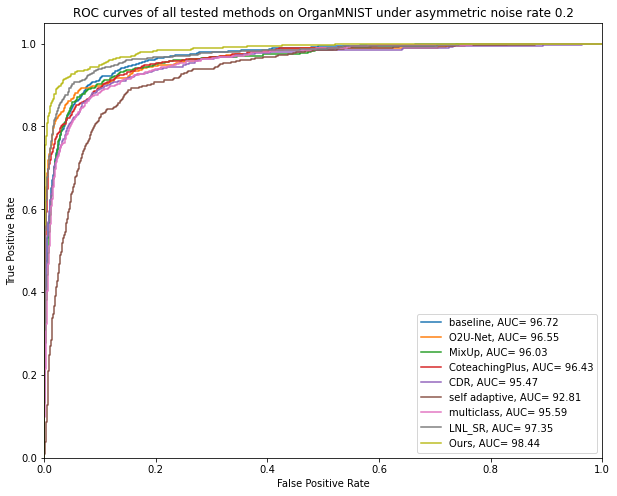}
	\caption{ROC curve and AUC of different methods on OrganMNIST3D dataset under asymmetric noise rate 0.2}
	\label{rocorgana}
\end{figure}

%ROC vessel
\begin{figure}[H]
	\centering
	\includegraphics[scale=0.42]{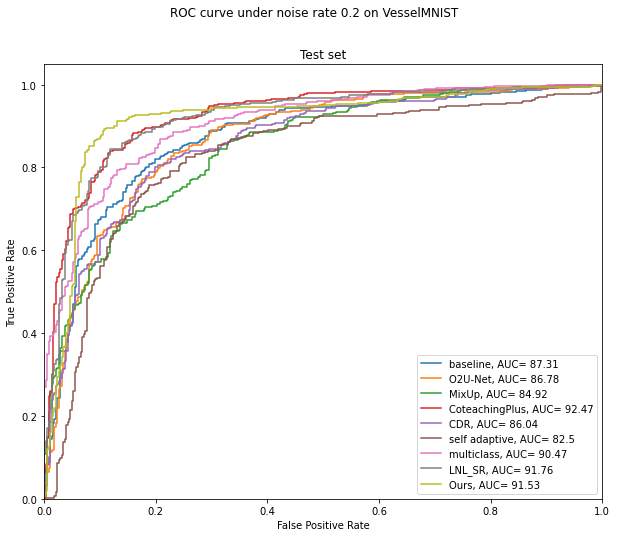}
	\caption{ROC curve and AUC of different methods on VesselMNIST3D dataset under noise rate 0.2}
	\label{rocvessels}
\end{figure}

\end{document}